 \documentclass[%
  reprint,
  amsmath,amssymb,
  aps,
  prb,
 ]{revtex4-1}

\usepackage{graphicx}
\usepackage{dcolumn}
\usepackage{bm}
\usepackage{color}
\usepackage{times}
\usepackage{amsmath}
\usepackage{float}
\usepackage[toc,page]{appendix}
\usepackage{textcomp}
\usepackage{lipsum}
\usepackage{graphicx}
\usepackage{bm}
\usepackage{color}
\usepackage{times}
\usepackage{amsmath}
\usepackage{float}
\usepackage{hyperref}
\usepackage{ulem} 
\usepackage[dvipsnames]{xcolor}

\newcommand{\beginappendix}{%
	\setcounter{table}{0}
	\renewcommand{\thetable}{A\arabic{table}}%
	\setcounter{figure}{0}
	\renewcommand{\thefigure}{A\arabic{figure}}%
	\setcounter{equation}{0}
	\renewcommand{\theequation}{A\arabic{equation}}%
}

\begin{document}


\title{Harnessing re-programmable phase transitions to control the propagation of sound waves}

\author{Audrey A. Watkins, Austin Eichelberg} 
\author{Osama R. Bilal} \email{osama.bilal@uconn.edu}

\affiliation{Department of Mechanical engineering, University of Connecticut, Storrs, CT 06269, USA}
\date{\today}
\begin{abstract}
Metamaterials can enable peculiar static and dynamic behavior (such as  negative effective mass density, dynamical stiffness, and Poisson's ratio) due to their geometry rather than their chemical composition. The geometry of these metamaterials can be thought of as the phase of the material, which is usually fixed once the material is fabricated. While there exist many theoretical and numerical studies of metamaterials that can change phase, or re-program, experimental realizations remain limited  due to challenges in manufacturability, the destructive nature of the re-programming and inherent non-linearities. Through a combination of analytical, numerical and experimental analyses, we utilize tunable, self-assembled, nonlinear magnetic lattices to realize metamaterials with reversible phase transitions. Our metamaterials are composed of free-floating disks, with embedded permanent magnets, confined within magnetic boundaries. We exploit the non-destructive nature of the adjustable magnetic boundaries to create a set of re-programmable metamaterials to control the propagation of sound waves. Furthermore, we demonstrate a robust, real-time tunable wave filter at ultra-low frequencies. Our findings can expand the metamaterials horizon into functional and tunable devices.
\end{abstract}


\maketitle

Metamaterials, often composed of repeated, identical unit cells, are structures with unique properties such as negative Poisson's ratios,\cite{lakes1987foam} vanishing shear modulus,\cite{milton1995elasticity,kadic2012practicability,buckmann2014elasto} negative stiffness,\cite{lakes2001extreme,nicolaou2012mechanical} and polar elasticity.\cite{bilal2017intrinsically}
Such metamaterials could have potential applications in wave control,\cite{bertoldi2008wave,wang2013effects,shan2014harnessing,shim2015harnessing,raney2016stable,bilal2017bistable,bilal2017reprogrammable,li2012switching,airoldi2011design} and acoustic cloaking.\cite{cummer2007one} The unique properties of these materials stem from their geometric tessellations, rather than their chemical compositions. These tessellations can be designed such that an applied external stimulus can change the geometric arrangements of the unit cells, giving rise to new domains, or phases, within the material.\cite{jin2020guided} Each of these phases resembles a stable configuration of the material. The stability of the metamaterial can be engineered such that multiple stable phases exist. \cite{shan2015multistable,chen2017harnessing,jiang2019bifurcation,zareei2020harnessing,schaeffer2015wave} The ability of metamaterials to transform, \textit{or be programmed}, into distinct stable phases correlates with the presence of multiple local minima in the metamaterial energy landscape.\cite{jin2020guided,shan2015multistable,benichou2013structures, restrepo2015phase,frazier2017atomimetic,zhang2019programmable,chen2019phase,zhang2019energy,yasuda2020transition,yang2016phase} One avenue to engineer phase transitioning metamaterials with multiple energy minima is to capitalize on soft materials and their ability to withstand large deformations without undergoing structural damage.\cite{li2012advances,li2012mechanics} This could facilitate the design of metamaterials with reversible phase transitions.
\medskip

Reversible transitions between phases, or energy minima, can be thought of as re-programmbility of a material. Re-programmable metamaterials offer more flexibility and adaptability in applications such as controlling electromagnetic \cite{ma2014acoustic,yang2014all,bao2018design,li2012flat,cui2016information,arbabi2017planar, lee2012reversibly} and acoustic waves. \cite{janke2005applications} While there exist many theoretical and numerical studies of programmable metamaterials, experimental realizations remain limited \cite{wang2020tunable} due to challenges in manufacturability,\cite{meloni2021engineering} the destructive nature of the re-programming, and the inherent non-linearity\cite{lazarov2007low,nadkarni2014dynamics,moleron2014solitary,narisetti2010perturbation,fronk2017higher,zaera2018propagation} within phase transitions. To overcome the damaging nature of re-programmability and the common challenge of manufacturing, we utilize easy-to-create, tunable, nonlinear metamaterials capable of filtering sound waves through non-destructive, re-programmable, and reversible phase transitions. 

\medskip

\begin{figure}[b]
	\begin{center}
		\includegraphics[scale=1.03]{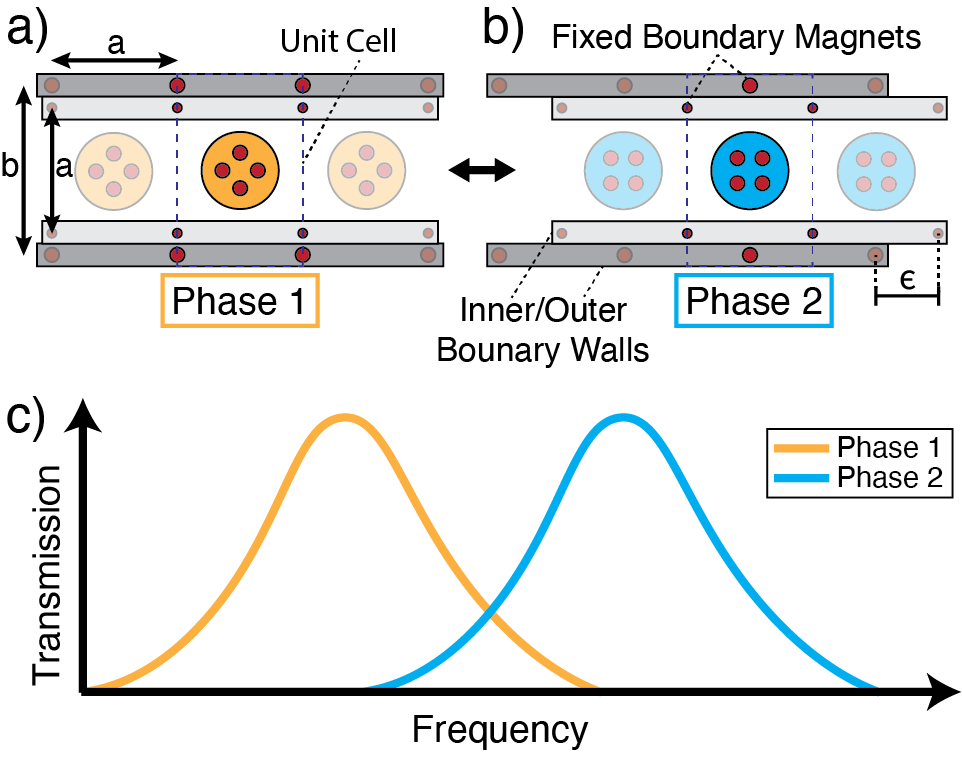}
		\caption{\label{fig:concept} \textbf{Concept.} Schematic of the metamaterials' unit cell \textcolor{black}{(opaque)} used in the experiments. The free floating disk is self-aligned in a) ``plus sign" or b) ``square" orientation based on the relative position of the boundary walls. c) Programmable transmission spectrum of the metamaterial when in phase 1 (orange) and phase 2 (blue). }
	\end{center}
\end{figure}


In this paper, we utilize tunable, self-assembled lattices of repelling magnets to create metamaterials that can control the propagation of sound waves by means of phase transitions. We create our lattices by confining self-aligned, free-floating disks with embedded permanent magnets to a fixed boundary with a \textit{pre-engineered} magnetic field. The magnetic repulsive forces between the boundary magnets and the free-floating disks cause each disk to self-align and rest within a magnetic potential well. Each magnetic potential well and its respective floating disk act as a single building block, or a unit cell, of the lattice. These identical unit cells repeat in space, forming a one-dimensional metamaterial with unique wave propagation characteristics. We design the geometry of the confining boundary such that the precise tuning of the magnetic field controls the alignment of the disks. The relative positions and orientation angles between the disks define the corresponding metamaterial phase. By changing the boundary geometry and therefore the resulting magnetic couplings within the unit cell, the transmittable frequencies through the metamaterial can be programmed.
\medskip

We create a unit cell that consists of a single disk bound by two sets of adjustable boundary walls (Fig. \ref{fig:concept}). The \textit{outer} set of the boundary walls contains four fixed cylindrical magnets, each with a diameter of 3 $mm$ and height of 3 $mm$, spaced by a distance $a$ horizontally and $b$ vertically. The distances $a$ and $b$ define the size of the unit cell. The \textit{inner} set of the boundary walls contains four fixed cylindrical magnets, each with a diameter of 2 $mm$ and height of 1 $mm$, spaced by the same distance $a$, both horizontally and vertically (Fig. \ref{fig:concept}). The fixed magnets form a magnetic potential well within the boundary, confining a floating disk that is embedded with multiple magnets (e.g., four magnets) identical to those in the outer boundary walls (i.e., 3 $mm$ height and diameter). Both \textit{inner} and \textit{outer} boundaries can slide against each other in a parallel fashion. When the magnets in both boundaries are vertically aligned with each other, the resultant magnetic field causes the free-floating disks to self-align in their unit cells with an orientation resembling a ``plus sign." This orientation is phase 1 of our metamaterial (orange disk in figure \ref{fig:concept} a). By sliding the set of inner boundary walls by $\epsilon = a/2$, such that the inner boundary magnets are halfway between the outer boundary magnets, the resultant magnetic field causes each disk to reorient, resembling a ``square shape." This orientation is phase  2 of our metamaterial (blue disk in figure \ref{fig:concept} b). Each metamaterial phase has its own wave transmission properties due to the different \textcolor{black}{magnetic} couplings between the boundaries and \textcolor{black}{the adjacent unit cells'} disks\textcolor{black}{, allowing for transmission by means of disk oscillation} (Fig. \ref{fig:concept} c). The boundary walls can be adjusted an infinite number of times to reorient the floating disks from phase 1 to phase 2 orientation and vice versa, allowing for the \textit{re-programmability} of the metamaterials.

\begin{figure}
\begin{center}
\includegraphics[scale=1.04]{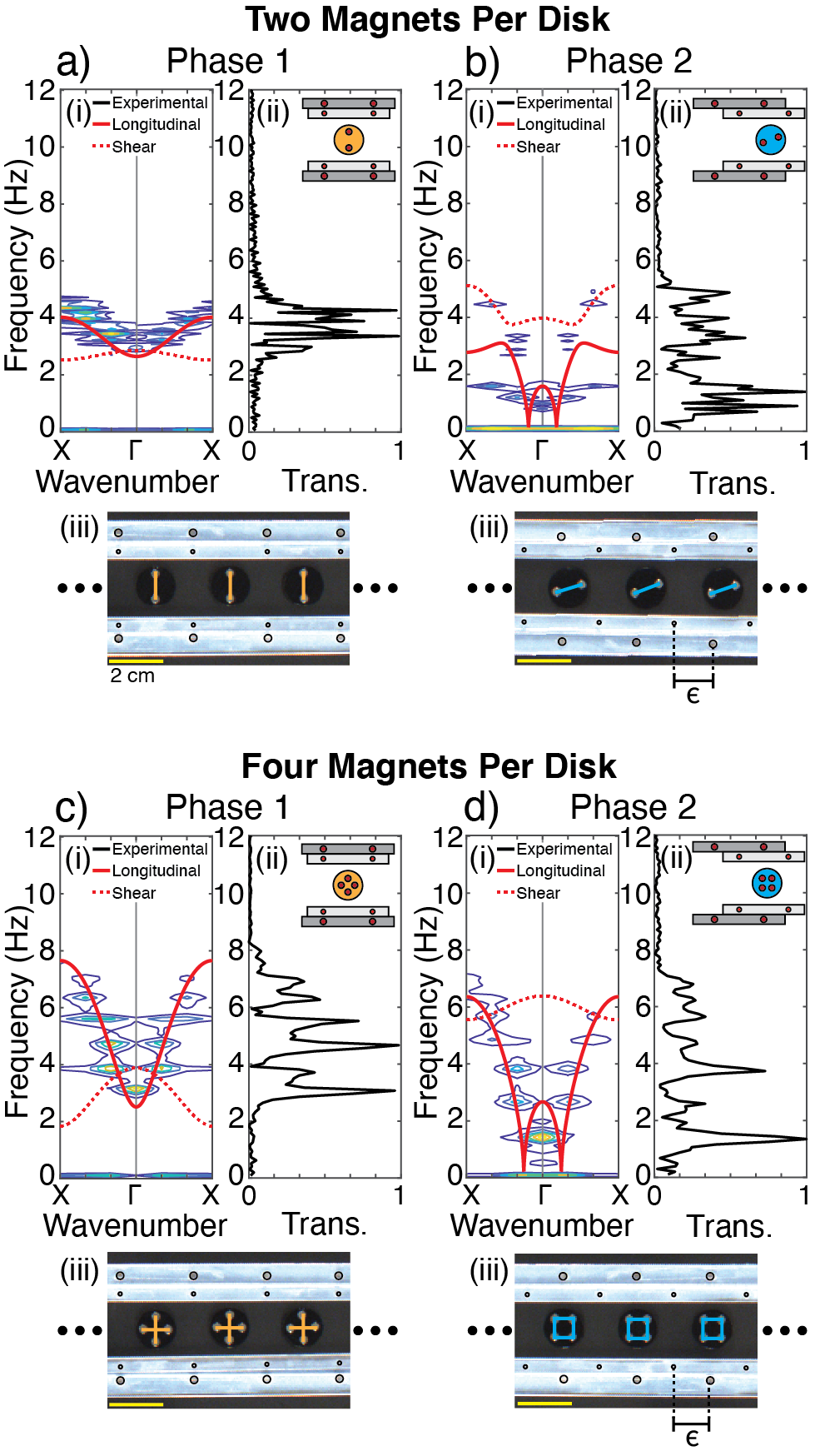}
\caption{\label{fig:channel_exp_disp_fft} \textbf{Programmable transmission spectrum.} Analytical dispersion curves and experimental transmission ranges for two magnet per disk in a) phase 1 and b) phase 2 orientation. Four magnet per disk behavior in the c) phase 1 and d) phase 2 orientation. (a-d)(i) The analytical (red) and experimental (contour) dispersion curves, (ii) the normalized transmission of a central disk and the unit cell (inset), and (iii) a photo of three unit cells within the metamaterial of the straight channel boundary in the phase 1 or phase 2 orientation (i.e., $\epsilon = 0$ or $\epsilon = a/2$, respectively).}
\end{center}
\end{figure}

We consider two types of disks (1) disks with two embedded magnets, and  (2) disks with four embedded magnets. The lattices composed of these disk types are stable in two topologically distinct  phases.  In the case of two embedded magnets per disk, the lowest energy states take place when the disks self-align in an in-plane ``vertical" orientation (phase 1) (Fig. \ref{fig:channel_exp_disp_fft} a(ii) inset) or in a ``slanted horizontal" orientation (phase 2) (Fig. \ref{fig:channel_exp_disp_fft} b(ii) inset). In the case of four embedded magnets per disk, the lowest energy states exist when the disks are in a ``plus sign" orientation (phase 1) (Fig. \ref{fig:channel_exp_disp_fft} c(ii) inset) or in a ``square" orientation (phase 2) (Fig. \ref{fig:channel_exp_disp_fft} d(ii) inset). The transition between the two low energy states (i.e., the two different phases) can be achieved by simple sliding of the set of inner boundary walls, while the set of outer boundary walls is fixed.

To analyze our metamaterials, we consider the unit cell as the basic building block of each lattice. We use Bloch's theorem \cite{bloch1929quantenmechanik} to calculate the dispersion curves of our metamaterials  by assuming that the unit cell repeats infinitely in space. The dispersion curves represent a correlation between frequency and wavenumber, revealing the expected transmission regions within the frequency spectrum. The considered unit cell, for each metamaterial phase, is composed of the free-floating disk, the fixed boundary magnets, and the neighboring free-floating disks on each side (i.e., nearest neighbor interactions (Fig. \ref{fig:supp_straight_channel_vectors})). The dispersion relation can be written as an eigenvalue problem:
\begin{equation}
    [-\omega^2\textbf{M}+\textbf{K}(\boldsymbol\kappa)] \boldsymbol{\phi} = 0,
    \label{eqn:eigen}
\end{equation}
where $\omega$ is the frequency, $\boldsymbol {M} = \begin{bmatrix}
m_i & 0\\
0 & m_i
\end{bmatrix}$ is the mass matrix and $m_i$ is the mass of the disk with $i$ embedded magnets, where $i=2$ or $4$ magnets. $\boldsymbol{K(\kappa)}$ is the stiffness matrix (Eq. \ref{eqn:stiff}) as a function of the wavenumber, $\kappa$. $\boldsymbol\phi = [u~v]^{T}$ is the Bloch displacement vector in the $x$ and $y$ directions, respectively. 
\medskip

We compute the analytical dispersion curves for a unit cell with dimensions $a = 27~mm$, $b =37~mm$, and $m_2=0.96$ grams (two embedded magnets per disk) or $m_4=1.235$ grams (four embedded magnets per disk) (Fig. \ref{fig:channel_exp_disp_fft}). The generated dispersion curves account for both the longitudinal and shear modes\textcolor{black}{, which can also be described as the parallel polarization and perpendicular polarization of the metamaterial, respectively}. The calculated dispersion curves for each phase are distinct from one another for both disk types. Notably, in the phase 1 orientation for both types of disks (Fig. \ref{fig:channel_exp_disp_fft} a(i), c(i)), neither of the transmission bands exist below a frequency of 1.8 Hz. In the phase 2 orientation for both types of disks (Fig. \ref{fig:channel_exp_disp_fft} b(i), d(i)), the emergent pass bands are almost grounded to a frequency of 0 Hz, indicating lower frequency waves are transmittable through this phase. Therefore, changing between phases for either disk type affects the lowest transmittable frequencies. The calculated shear and longitudinal dispersion curves intersect in all cases except for two magnets per disk in the phase 2 orientation, where the shear and longitudinal dispersion curves are separated by a stop band (Fig. \ref{fig:channel_exp_disp_fft} b(i)). In the case of two magnets per disk in phase 1, the calculated transmission band spans 1.5 Hz with an upper cutoff frequency of 4 Hz (Fig. \ref{fig:channel_exp_disp_fft} a(i)). In contrast, in the case of four magnets per disk in phase 1, the calculated transmission band spans 5.8 Hz ($\sim$ 4 times wider) with an upper cutoff frequency of 7.6 Hz (Fig. \ref{fig:channel_exp_disp_fft} c(i)).

To experimentally validate the different phases of our metamaterials, we realize two sets of adjustable boundary walls resembling those in figure \ref{fig:concept}. The boundaries are cut from an acrylic glass sheet using a laser cutter (Full-spectrum 24 pro-series). We fabricate the floating disks, with a diameter $d = 15~mm$, using a 3D printer (Connex 500). All embedded magnets in both the boundary walls and the disks have the north pole facing upwards, replicating the behavior of monopoles in plane. We place the boundary walls atop an air bearing table (New way S1030002). We adhere a glass slide to the bottom side of each disk to minimize friction, and place the disks within the sets of boundary walls in random orientations. Then, we activate the air bearing table and the resulting laminar flow beneath the disks creates a near frictionless surface. \textcolor{black}{The laminar layer of airflow beneath the disks has a fly height of 5 microns (about three orders of magnitude smaller than the longitudinal and shear displacements of the disks), allowing us to consider the flexural motion negligible.} The disks move freely within their magnetic potential wells, self-aligning into their lowest potential energy locations and orientations. The observed low energy states, or phases, correspond to the analyzed lattices in figure \ref{fig:channel_exp_disp_fft} (a-d)(iii). The re-programmability of the lattices, or the reversible transition between these different phases, is achieved by sliding the boundaries [see supplementary video 1]. 
\medskip

To experimentally verify the analytically computed dispersion curves, we test the dynamical characteristics of the different metamaterial phases once they self-assemble (Fig. \ref{fig:channel_exp_disp_fft} (a-d)(iii)). We excite the right-most disk in each channel longitudinally (i.e., along the direction of periodicity) with a chirp signal between 0.2 Hz and 35 Hz using a mechanical shaker (Br\"{u}el and Kj\ae r 4180) and a function generator (Keysight Technologies 33512B). Throughout the duration of the chirp signal, we capture the motion of each disk using a computer vision camera (Blackfly S USB3). We then post-process the images using the digital image correlation software (DICe) and record the displacement of each disk in time. The mechanical waves propagate through the lattice in the form of an oscillatory motion of the free-floating disks. We analyze the x-displacement (the direction of excitation) profiles of the floating disks using 1D and 2D fast Fourier transform (2D-FFT). The 1D-FFT of a central disk represents the transmission of the wave through the lattice at different frequencies  (Fig. \ref{fig:channel_exp_disp_fft} (a-d)(ii)). The resulting 2D-FFT is superimposed as a contour map over the analytical dispersion curves for each metamaterial (Fig. \ref{fig:channel_exp_disp_fft} (a-d)(i)). 
\medskip

The experimentally observed transmission ranges match well with the analytical results. In addition, the transmission ranges obtained numerically through a finite simulation using the Verlet method \cite{press2007numerical} is in agreement with both the analytical and experimental data for phase 1 (Fig. \ref{fig:channel_exp_disp_fft_with_numerics}). As noted above, the increase in the number of embedded magnets in the disks causes both the upper frequency cutoff values and the widths of transmittable frequencies to increase. Similarly, the change in phase orientation from phase 1 to phase 2 lowers the range of transmittable frequencies. It is worth mentioning that the shear mode is not excited in our experiments. However, we note the presence of shear motion in the case of two magnets per disk in phase 2, due to the coupling between the longitudinal and shear modes. The mode coupling is a consequence of the ``slanted horizontal" (as opposed to perfectly horizontal) orientation of the disks in this phase, which we show numerically in figure \ref{fig:shear_presence_two_mag}. We confirm the nature of the propagating modes in each metamaterial phase by applying a pure shear excitation (i.e., perpendicular to the direction of periodicity) (Fig. \ref{fig:shear_channel_exp_disp_fft}).
\medskip

During the experiments, we observed a boundary effect on the left-most and right-most floating disks. Upon shifting the boundary walls from phase 1 to phase 2 and vice versa, the left-most and right-most floating disks did not always autonomously change phase with the rest of the disks (i.e., sometimes they remained in their initial phase orientations). These inconsistencies can cause non-uniform re-programmability that has the potential to cause discrepancies between the finite experimental and infinite analytical models in more complex and large-scale systems. 
\medskip

\begin{figure}
	\begin{center}
		\includegraphics[scale=1.04]{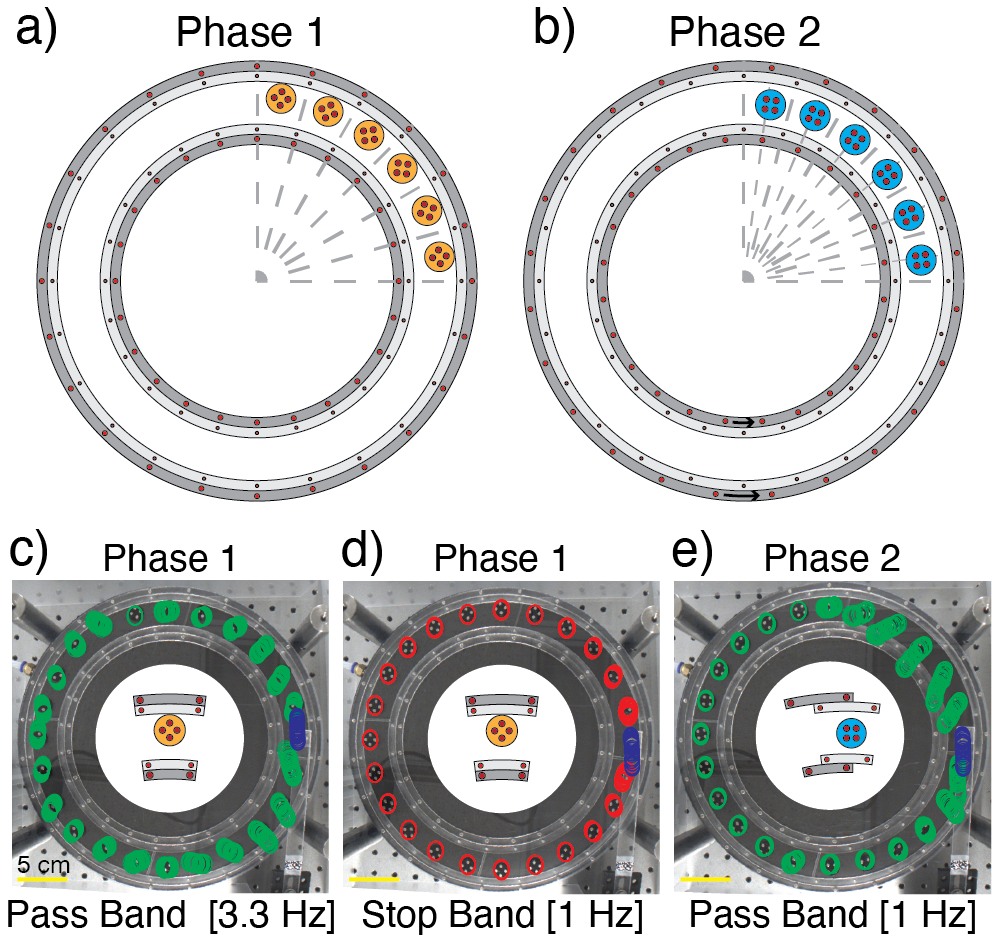}
		\caption{\label{fig:circle_disk_location_tracking} \textbf{Circular metamaterial dynamics.} Schematic of the circular channel filled with six of the twenty-four floating disks embedded with four magnets in each. Boundary alignments in the a) phase 1 and b) phase 2 orientation. The displacements, scaled by a factor of 70, of each disk (green or red) and the excitation source (blue) within the circular boundary in the phase 1 orientation when excited at c) 3.3 Hz and d) 1 Hz and e) in the phase 2 orientation when excited at 1 Hz.}
	\end{center}
\end{figure}

To overcome the issues that could arise from the boundary effects, we design a re-programmable metamaterial that has no left or right end: an infinite circular channel.  Similar to the inner and outer boundary walls in the straight channel, we cut a boundary consisting of adjustable concentric circles from acrylic glass. We embed the same 3 $mm$ x 3 $mm$ magnets in the outer boundary walls and the 2 $mm$ x 1 $mm$ magnets in the inner boundary walls, creating 24 magnetic potential wells (i.e., 24 unit cells). When the floating disks are placed within the unit cells, they self-align in a manner similar to the straight channel in either the phase 1 (Fig. \ref{fig:circle_disk_location_tracking} a) or phase 2 (Fig. \ref{fig:circle_disk_location_tracking} b) orientation, depending on the relative alignment between the boundary walls. We note that in the phase 1 orientation the floating disks lay against the outer circular boundary ring.
\medskip

To analytically model the unit cells within the circular boundary, we utilize the same Bloch analysis equations (Eqs. \ref{eqn:eigen}, \ref{eqn:stiff}) while accounting for the curved boundary and the new equilibrium positions of the disks (Fig. \ref{fig:supp_circular_channel_vectors}). To experimentally verify the dynamical characteristics of the different metamaterial phases, we excite the metamaterial at the right vertical tangent to the circular boundary. Similar to the straight channel, we post-process the displacement of all the disks in time using the DICe software. The analytical dispersion, experimental transmission, and experimental dispersion are plotted in figure \ref{fig:circle_exp_disp_fft_updated}. 
\medskip

To demonstrate the programmability of our system in the infinitely periodic circular channel, we excite the metamaterial in phase 1 at two different frequencies corresponding to a pass band (at 3.3 Hz) and a stop band (at 1 Hz). Then, we induce a phase transition within our metamaterial by rotating the outer boundary walls by $7.5^{\circ}$ (Fig. \ref{fig:circle_disk_location_tracking} a, b). As a result, all disks change alignment from the ``plus sign" orientation into the ``square" orientation (i.e., phase 2). Consequently, the dynamical characteristics of the metamaterial change, and the frequency of 1 Hz is no longer within the stop band, but rather is a pass band frequency. To visualize the behavior of the metamaterials at different phases and excitation frequencies, we superimpose the displacement profiles of all the disks on the circular channel images (Fig. \ref{fig:circle_disk_location_tracking} c-e). The displacement of the excitation source is shown in blue. The displacements of all other 23 freely floating disks are shown in green (in case of pass band frequency) or red (in case of stop band frequency). In case of phase 1 excitation at 3.3 Hz, the displacement profiles show clear transmission of the wave through the metamaterial (Fig. \ref{fig:circle_disk_location_tracking} c). For the same metamaterial phase excited at 1 Hz, the displacement profiles indicate an exponential decay of the wave (Fig. \ref{fig:circle_disk_location_tracking} d). At the same excitation frequency of 1 Hz, after the metamaterial transitions into phase 2, the wave transmits through (Fig. \ref{fig:circle_disk_location_tracking} e).   
\medskip

\begin{figure}[b]
	\begin{center}
		\includegraphics[scale=1.02]{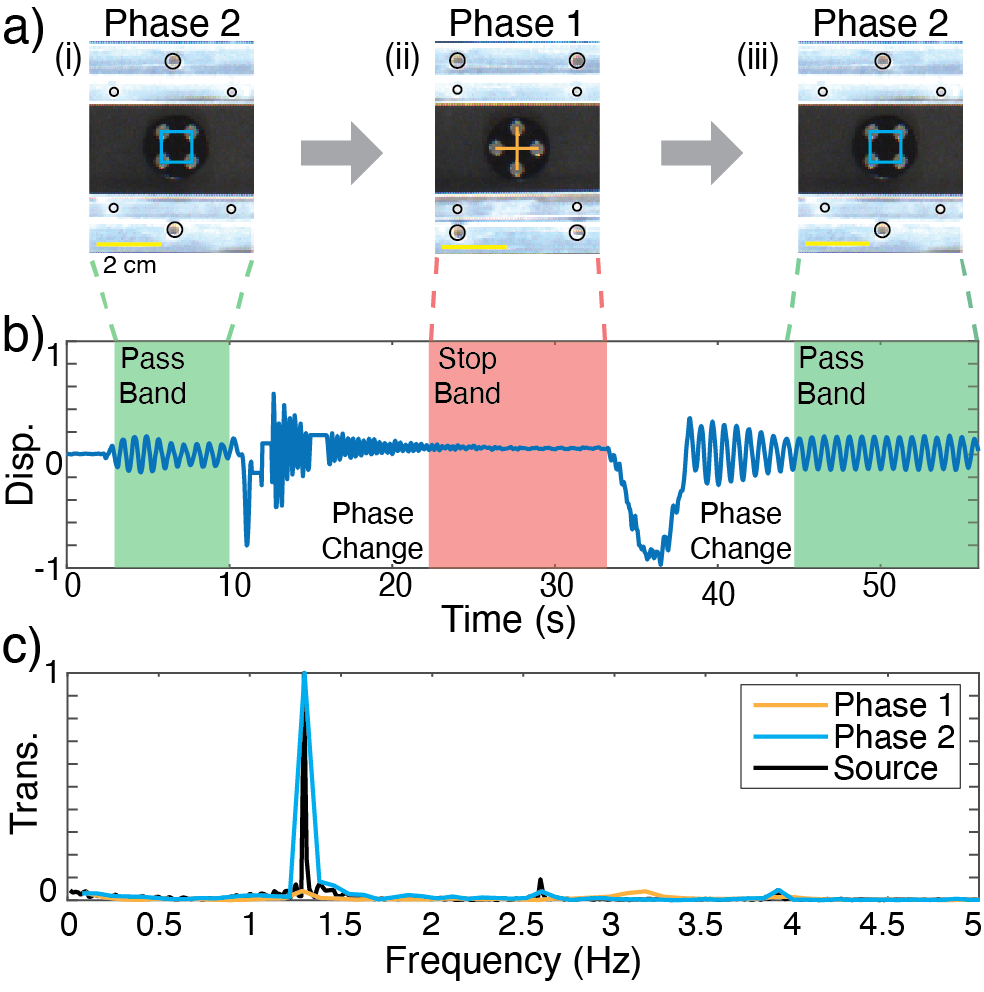}
		\caption{\label{fig:time_dependent_transmission} \textbf{Re-programmability in time.} a) Central unit cell of the straight channel metamaterial with four magnets embedded in each floating disk transitioning from the the (i) phase 2 to (ii) phase 1 back to (iii) phase 2 orientation. b) Displacement in time of the central disk as the metamaterial is excited at a frequency $f=1.3$ Hz, the green highlight corresponding to the phase 1 orientation and the red highlight corresponding to the phase 2 orientation. c) The 1D-FFT of the central disk in phase 1 (orange) and phase 2 (blue), and of the vibration source signal (black).}
	\end{center}
\end{figure}

Inducing a phase transition in both the straight and circular channels is reversible, re-programmable, and relatively simple.  We harness these re-programmable phase transitions as a real-time tunable wave filter. To validate the concept of real-time tunability, we consider a metamaterial composed of unit cells with four embedded magnets in each disk within the straight channel. We induce cyclic phase transitions in the metamaterial (Fig. \ref{fig:time_dependent_transmission} a) as we continuously excite it longitudinally at 1.3 Hz, corresponding to a pass band in phase 2 and a stop band in phase 1 (Fig. \ref{fig:channel_exp_disp_fft} c, d). We track the displacement of a central disk, while the lattice is being excited, through the different phase transitions (Fig. \ref{fig:time_dependent_transmission} b). At time $t=0$ seconds, the excitation is off and the metamaterial is in phase 2 (Fig. \ref{fig:time_dependent_transmission} a(i)). At $t=2$ seconds, the excitation source is on and the disk oscillates. At $t=11$ seconds, we start the phase transition into phase 1 (Fig. \ref{fig:time_dependent_transmission} a(ii)). Due to the adjustment of the boundary walls during the phase transition, the disks briefly register large displacements as they self-align into their phase 1 orientations. At $t=13$ seconds, the oscillatory motion of the disks begins to decay exponentially and the disks reach their phase 1 positions and orientations, approaching static equilibrium. At $t=33$ seconds, we start the phase transition back into phase 2 (Fig. \ref{fig:time_dependent_transmission} a(iii)). The disks briefly register large displacements during their shift back to the phase 2 orientation. At $t=38$ seconds, the disks reach their steady-state behavior and show a constant oscillatory motion, indicating the propagation of the sound waves through the metamaterial. To identify the main frequency components within the time signal, we apply 1D-FFT to the regions of attenuation (red) and transmission (green) (Fig. \ref{fig:time_dependent_transmission} c). We also apply 1D-FFT to the signal of the vibration source. The FFT shows clear transmission in the phase 2 orientation (blue) and almost zero transmission in the phase 1 orientation (orange). The time signal during the phase transitions demonstrates the behavior of a real-time tunable wave filter.
\medskip

We have utilized tunable, self-assembled, nonlinear magnetic lattices to create metamaterials with reversible phase transitions. We harness these controllable phase transitions to engineer the propagation of sound waves. Our metamaterials are composed of free-floating disks with embedded permanent magnets confined within magnetic boundaries. We exploit the non-destructive nature of the adjustable magnetic boundaries to create a set of re-programmable metamaterials within a straight channel. We further improve our straight channel to mitigate the boundary effects by considering an infinitely periodic circular arrangement of our metamaterials. Furthermore, we demonstrate a robust, real-time tunable wave filter at ultra-low frequencies. The applications for metamaterials with re-programmable phase transitions are extensive, including wave guiding, cloaking, and vibration control. Our findings can expand the capabilities of metamaterials used in functional and tunable devices.

%
\newpage
\bibliographystyle{MSP}
\bibliography{Phase}

\newpage

\begin{appendices}
\beginappendix

\section{Straight Boundary Analysis}

\begin{figure}
\begin{center}
\includegraphics[scale=1]{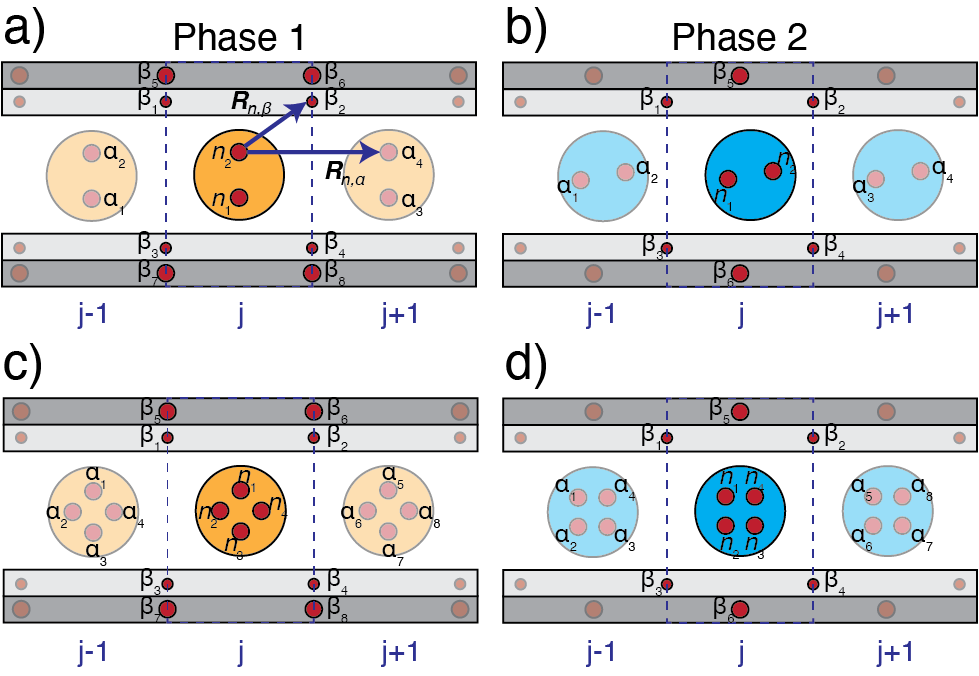}
\end{center}
\caption{\textbf{Straight channel unit force vectors.} Labeling and representation of the unit force vectors within a unit cell, $j$ (opaque). a) Example unit vector between disk $j$ magnets ($n$), boundary magnets ($\beta$), and the adjacent disks' embedded magnets $\alpha$ (transparent) used in the stiffness matrix. Labeling and indexing of two magnets per disk b) phase 2 and four magnets per disk c) phase 1 and d) phase 2.}
\label{fig:supp_straight_channel_vectors}
\end{figure}

In our infinite model analysis, i.e., the dispersion curve calculations, we consider the nearest neighbor interactions between the disks. We consider a periodically repeating unit cell with disk index $j$. Each disk can have two or four embedded magnets. The stiffness matrix in the utilized model (Eq. \ref{eqn:stiff}) accounts for the repulsion forces between each embedded magnet ($n_i$), boundary magnet ($\beta_{ii}$) in the same unit cell, and the disk magnet ($\alpha_{iii}$) in the adjacent unit cells $j+1$ and $j-1$. The stiffness matrix, $\boldsymbol{K(\kappa)}$, consists of the static repulsive forces between the disk magnets and the boundary magnets in the lowest potential energy locations (Fig. \ref{fig:supp_straight_channel_vectors}) and can be written as \cite{watkinseichelbergbilal2021frustration}:

\begin{multline}
\mathbf{\textbf{K}}=\sum_{n=1}^{N} \Biggl( 
\sum_{\alpha=1}^{2N} \biggl[ f_{,d}(d_{n,\alpha})_{l}\textbf{e}_{n,\alpha}\otimes \textbf{e}_{n,\alpha}\left [ cos(\kappa \cdot \textbf{R}_{n,\alpha}) 
-1 \right]\\ + \frac{f(d_{n,\alpha})_{l}}{(d_{n,\alpha})}(\textbf{I}-\textbf{e}_{n,\alpha}\otimes \textbf{e}_{n,\alpha})\left [ cos(\kappa\cdot \textbf{R}_{n,\alpha})-1 \right ] \biggr]\\ 
- \sum_{\beta=1}^{4}\biggl[ f_{,d}(d_{n,\beta})_{s}\textbf{e}_{n,\beta}\otimes \textbf{e}_{n,\beta} + \frac{f(d_{n,\beta})_{s}}{(d_{n,\beta})}(\textbf{I}-\textbf{e}_{n,\beta}\otimes \textbf{e}_{n,\beta})\biggr]\\
- \sum_{\beta=5}^{\Gamma}\biggl[ f_{,d}(d_{n,\beta})_{l}\textbf{e}_{n,\beta}\otimes \textbf{e}_{n,\beta} + \frac{f(d_{n,\beta})_{l}}{(d_{n,\beta})}(\textbf{I}-\textbf{e}_{n,\beta}\otimes \textbf{e}_{n,\beta})\biggr] \Biggr)
\label{eqn:stiff}
\end{multline}

where $N$ is the number of magnets embedded in a single floating disk, $n$ is the index of the magnets in unit cell $j$, $\alpha$ is the index of the magnets in the nearest neighboring unit cells (e.g. $j+1$, $j-1$), $\otimes$ is the dyadic product, $\boldsymbol{R_{n,\alpha}}$ is the unit vector between magnet $n$ and $\alpha$, $\beta$ is the index of the boundary magnets, and $\Gamma$ is the number of boundary magnets. In the phase 1 orientation (i.e., when $\epsilon=0$), $\Gamma=8$ which accounts for the four inner boundary magnets and the four outer boundary magnets of the unit cell. In the phase 2 orientation (i.e., when $\epsilon=a/2$), $\Gamma=6$ which accounts for all four inner boundary magnets and two outer boundary magnets (positioned halfway between the inner boundary magnets) of the unit cell. We model the repulsive forces measured between the embedded disk magnets and the boundary magnets as an inverse power law $f(d)_{(.)} = A_{(.)}d^{\gamma_{(.)}}$. $A_{(.)} = 3\mu \beta ^2/4\pi$ represents the interaction constant between the disk magnets and the smaller inner boundary magnets ($A_s = 2.210 \times 10^{-10}$) or the larger outer boundary magnets ($A_l = 9.617 \times 10^{-11}$) where $\mu$ is the permeability of air and $\beta$ is the magnetic moment, $f_{,d}()$ is the first derivative of $f()$, $\gamma_s = -3.036$, and $\gamma_l = -4$. \cite{jiao2019distinctive,watkins2020demultiplexing}
\medskip

Figure \ref{fig:supp_straight_channel_vectors} depicts a schematic of the unit cell vectors that appear in the stiffness matrix in equation \ref{eqn:stiff} for the straight channel boundary. Unit cells in the phase 1 orientations (Fig. \ref{fig:supp_straight_channel_vectors} a, c) have eight fixed boundary magnets (i.e., $\Gamma=8$) while unit cells in the phase 2 orientations (Fig. \ref{fig:supp_straight_channel_vectors} b, d) have six fixed boundary magnets (i.e., $\Gamma=6$). The number of boundary magnets within the unit cell corresponds to the $\Gamma$ value used in the stiffness equation (Eq. \ref{eqn:stiff}). The index of each disk magnet is labeled. An example unit vector between disk magnet $n_i$, where $i = 2$, and boundary magnet $\beta_{ii}$, where $ii = 2$ and the same disk magnet $n_{2}$ and adjacent disk magnet $\alpha_{iii}$, where $iii = 4$ is shown in figure \ref{fig:supp_straight_channel_vectors} a. The coupling between disk magnets $n$ in unit cell $j$ and disk magnets $\alpha$ in unit cells $j+1$ and $j-1$ resembles the nearest neighbor interactions accounted for in the dispersion analysis. 
\medskip

\begin{figure}
\begin{center}
\includegraphics[scale=1.04]{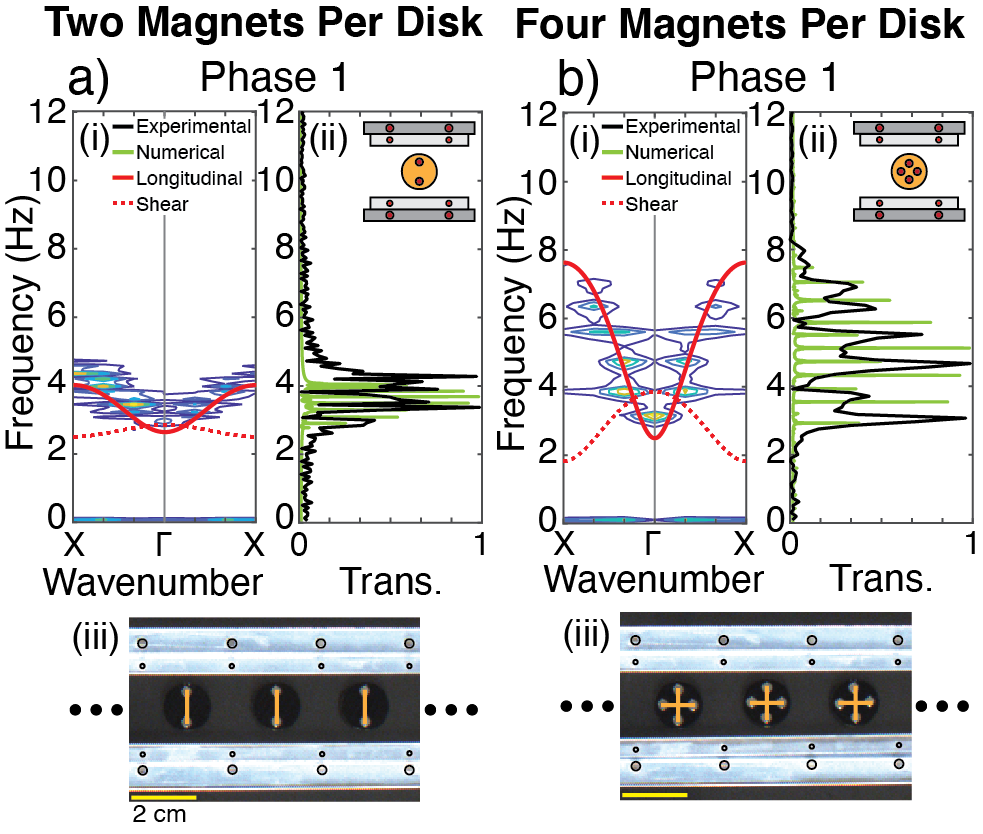}
\caption{\label{fig:channel_exp_disp_fft_with_numerics} \textbf{Longitudinal Excitation.} Numerical FRF of phase 1 metamaterial within the straight channel boundary with a) two magnets per disk and b) four magnets per disk. (a, b)(i) The analytical (red) and experimental 2D-FFT (contour map) dispersion curves. (a, b)(ii) The numerical FRFs (green) superimposed on the experimental transmission (black) through the metamaterial. (iii) photo of the self-assembled unit cells within the phase 1 metamaterials}
\end{center}
\end{figure}

To numerically validate the behavior of our metamaterials, we simulate the frequency response function (FRF) of a finite metamaterial array composed of 20 disks in phase 1 using the Verlet method. In the case of two magnets per disk (Fig. \ref{fig:channel_exp_disp_fft_with_numerics} a(ii) inset), the simulated FRF resembles the experimental 1D-FFT relatively well, accounting for a slight difference in the width of the range of transmission between the two plots (Fig. \ref{fig:channel_exp_disp_fft_with_numerics} a(ii)). The upper frequency value of the numerically simulated FRF is 4.1 Hz, whereas in the experimentally obtained 1D-FFT, it is 4.6 Hz. The lower frequency value of the numerically simulated FRF is 2.8 Hz, where as in the experimentally obtained 1D-FFT, it is 2.6 Hz.  In the case of four magnets per disk (Fig. \ref{fig:channel_exp_disp_fft_with_numerics} b(ii) inset), the numerical FRF and 1D-FFT match relatively well (Fig. \ref{fig:channel_exp_disp_fft_with_numerics} b). The upper frequency value of transmission of the numerically simulated FRF is 7.4 Hz, whereas in the 1D-FFT it is 8 Hz. The lower frequency value of transmission of the numerically simulated FRF is 2.9 Hz, whereas in the 1D-FFT it is 3.1 Hz. Overall, the numerically simulated FRFs are a good depiction of the behavior of the metamaterials in the phase 1 orientations when excited longitudinally. The slight discrepancy in the numerical FRFs and experimental ones is due to the variation in the size of the embedded magnets in the two different boundary sets, as these magnets do not have the same thickness and therefore are not in the same plane.
\medskip

\begin{figure}
\begin{center}
\includegraphics[scale=1]{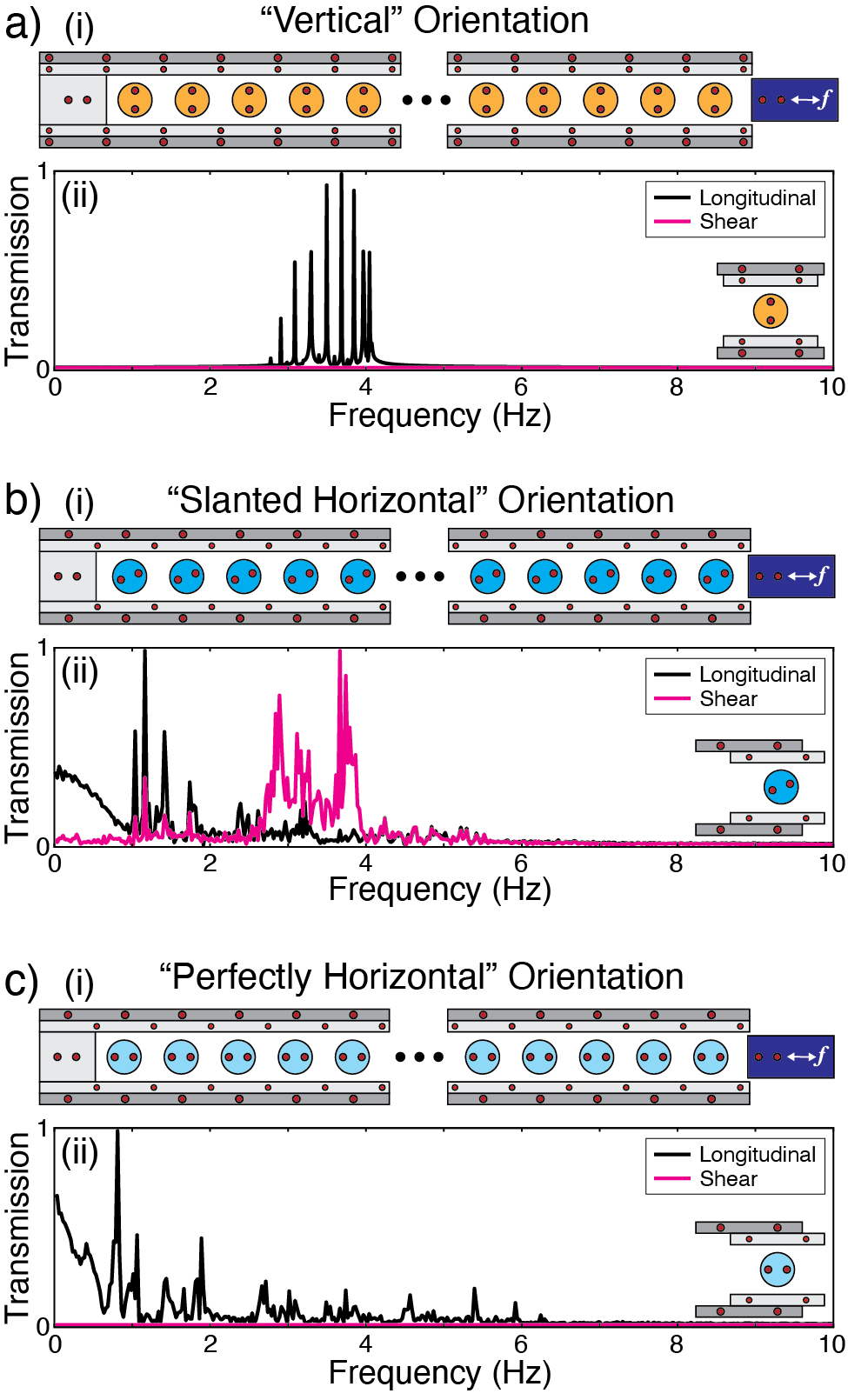}
\caption{\label{fig:shear_presence_two_mag} \textbf{Shear presence in longitudinal excitation.} Schematic of the two magnet per disk metamaterial in the a)(i) phase 1 ``vertical," b)(i) phase 2 ``slanted horizontal," and c)(i) phase 2 ``perfectly horizontal" orientation. (a-c)(ii)The numerically obtained FRF including both the longitudinal (black) and shear (pink) modes of transmission corresponding to the disk orientation.}
\end{center}
\end{figure}

Within the main text of the paper, we consider the longitudinal excitation (i.e., along the direction of periodicity) of the straight channel (Fig. \ref{fig:channel_exp_disp_fft}). This longitudinal excitation causes the displacement profiles of the disks to correspond with the longitudinal dispersion curves only, as long as the dispersion modes are not coupled (Fig. \ref{fig:channel_exp_disp_fft} a, c, d). When the modes are coupled and the metamaterial is longitudinally excited, shear motion presents itself as well (Fig. \ref{fig:channel_exp_disp_fft} b). For instance, in the case of two magnets per disk in phase 2, the ``slanted horizontal" orientation of the floating disks causes the shear and longitudinal modes to couple. To validate our observation, we numerically simulate three different phases of the metamaterials with two embedded magnets per disk using the Verlet method (Fig. \ref{fig:shear_presence_two_mag}). We compare the longitudinal versus shear motion of (1) vertically aligned disks (Fig. \ref{fig:shear_presence_two_mag} a(i)), (2) ``slanted horizontal" disks  (Fig. \ref{fig:shear_presence_two_mag} b(i)), and (3) perfectly aligned horizontal disks (Fig. \ref{fig:shear_presence_two_mag} c(i)). We excite each metamaterial \textit{longitudinally} and plot the simulated longitudinal and shear displacements for each configuration (Fig. \ref{fig:shear_presence_two_mag} (a-c)(ii)). The longitudinal excitation causes shear motion to occur only when the floating disks are in the ``slanted horizontal" orientation. We note that the perfectly horizontal alignment of the disks is not physical and such phase is not stable for these unit cell dimensions, however, we simulate its response to longitudinal excitation to show that the presence of the shear mode only exists when the disks are aligned at an angle (i.e., ``slanted").
\medskip

\begin{figure}
\begin{center}
\includegraphics[scale=1.04]{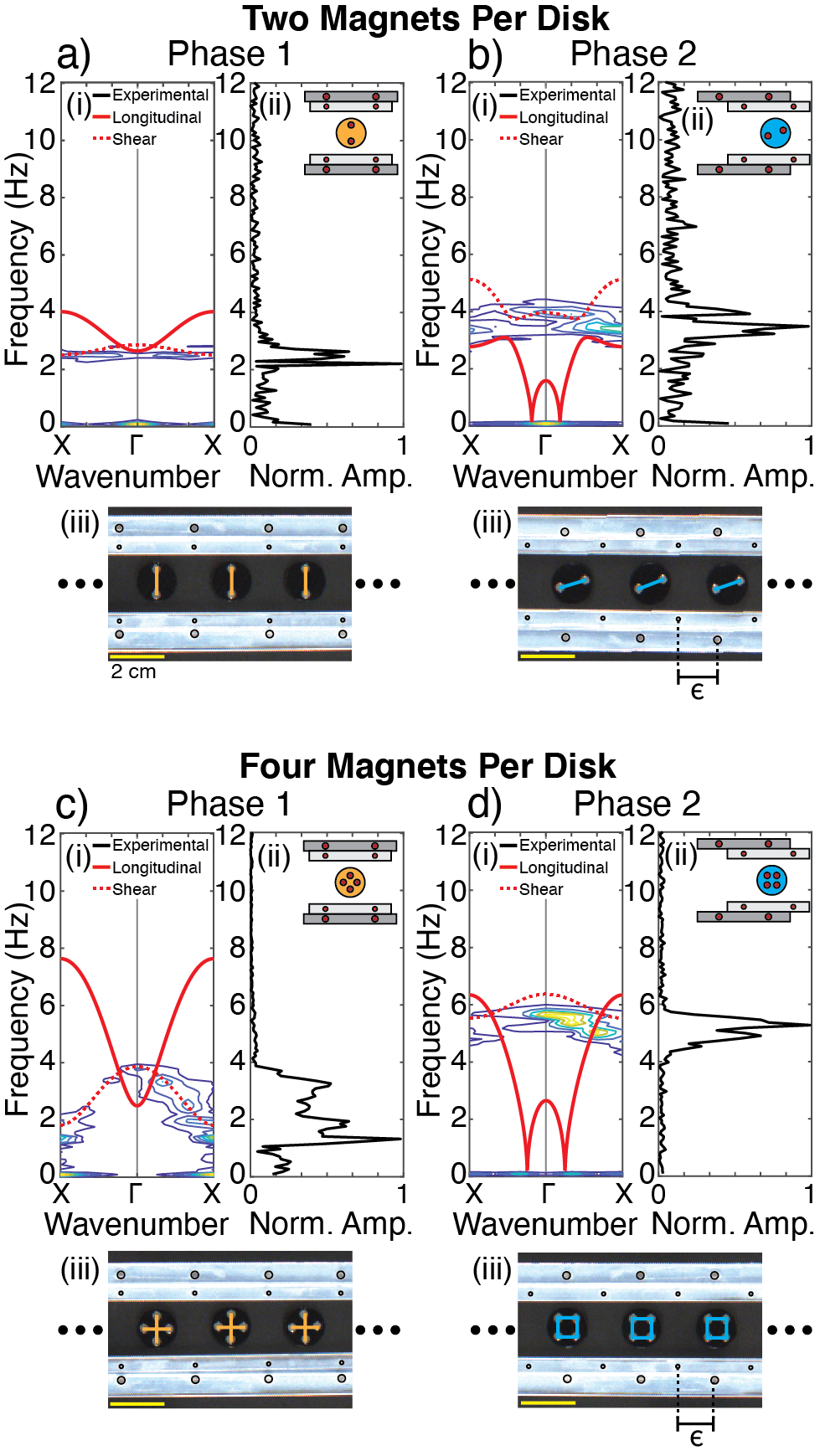}
\caption{\label{fig:shear_channel_exp_disp_fft} \textbf{Shear Excitation.} Two magnet per disk behavior when the shear mode is excited in the a) phase 1 and b) phase 2 orientation. Four magnet per disk behavior when the shear mode is excited in the c) phase 1 and d) phase 2 orientation. (a-d)(i) The analytical (red) and shear experimental (contour) dispersion curves, (ii) the normalized transmission of a central disk and the unit cell (inset), and (iii) a photo of three unit cells within the metamaterial of the straight channel boundary in the phase 1 or phase 2 orientation (i.e., $\epsilon = 0$ or $\epsilon = a/2$, respectively).}
\end{center}
\end{figure}

To experimentally confirm the nature of the propagating modes through our straight channel metamaterials (i.e., longitudinal versus shear), we excite the self-aligned lattices (Fig. \ref{fig:shear_channel_exp_disp_fft} (a-d)(iii)) with chirp signal perpendicular to the direction of periodicity.  The resulting motions of the floating disks are tracked using DICe, and 1D- and 2D-FFTs are used to post-process the y-displacements of the disks (Fig. \ref{fig:shear_channel_exp_disp_fft}). In the case of two magnets per disk in the phase 1 orientation (Fig. \ref{fig:shear_channel_exp_disp_fft} a), the analytical shear dispersion curve (Fig. \ref{fig:shear_channel_exp_disp_fft} a(i)) spans from 2.5 Hz to 2.9 Hz and the transmission (Fig. \ref{fig:shear_channel_exp_disp_fft} a(ii)) spans from 2.2 Hz to 2.8 Hz. In the case of two magnets per disk in the phase 2 orientation (Fig. \ref{fig:shear_channel_exp_disp_fft} b), the analytical shear dispersion curve (Fig. \ref{fig:shear_channel_exp_disp_fft} b(i)) spans from 3.8 Hz to 5.1 Hz and the transmission (Fig. \ref{fig:shear_channel_exp_disp_fft} b(ii)) spans from 2.9 Hz to 4.3 Hz. The transmission pass band that is present in phase 2 of two magnets per disk is shifted slightly lower than the shear dispersion curve, but the width of frequency values transmitted is very similar, 1.3 Hz versus 1.4 Hz for the analytical shear (Fig. \ref{fig:shear_channel_exp_disp_fft} b(i) (red)) and experimental data (Fig. \ref{fig:shear_channel_exp_disp_fft} b(ii)), respectively. In the case of four magnets per disk in the phase 1 orientation (Fig. \ref{fig:shear_channel_exp_disp_fft} c), the analytical shear dispersion curve (Fig. \ref{fig:shear_channel_exp_disp_fft} c(i)) spans from 1.8 Hz to 3.9 Hz and the transmission (Fig. \ref{fig:shear_channel_exp_disp_fft} c(ii)) spans from 1.3 Hz to 3.7 Hz. In the case of four magnets per disk in the phase 2 orientation (Fig. \ref{fig:shear_channel_exp_disp_fft} d), the analytical shear dispersion curve (Fig. \ref{fig:shear_channel_exp_disp_fft} d(i)) spans from 5.5 Hz to 6.4 Hz and the transmission (Fig. \ref{fig:shear_channel_exp_disp_fft} d(ii)) spans from 4.5 Hz to 5.7 Hz. In both phases with four magnets per disk, the analytical shear dispersion curves and transmission match relatively well, noting a slight downward shift of the transmission pass band.
\medskip

\textbf{Supplementary Video 1:} 
This video shows the re-programmability and reversibility of the straight metamaterial with four magnets embedded in each disk. First, we align the boundary walls such that the metamaterial is in phase 1 (each disk resembles a ``plus sign"). Then, we initiate the phase transition such that the disks reorient themselves and the metamaterial is in phase 2 (each disk resembles a ``square"). Lastly, we initiate another phase transition and the metamaterial returns to phase 1.
\medskip

\textbf{Supplementary Video 2:} 
This video shows the re-programmability and reversibility of the metamaterial within the circular boundary with four magnets embedded in each disk. First, we align the boundary walls such that the metamaterial is in phase 1 (each disk resembles a ``plus sign"). Then, we initiate the phase transition such that the disks reorient themselves and the metamaterial is in phase 2 (each disk resembles a ``square"). Lastly, we initiate another phase transition and the metamaterial returns to phase 1.
\medskip

\section{Circular boundary analysis}

\begin{figure}
\begin{center}
\includegraphics[scale=1.03]{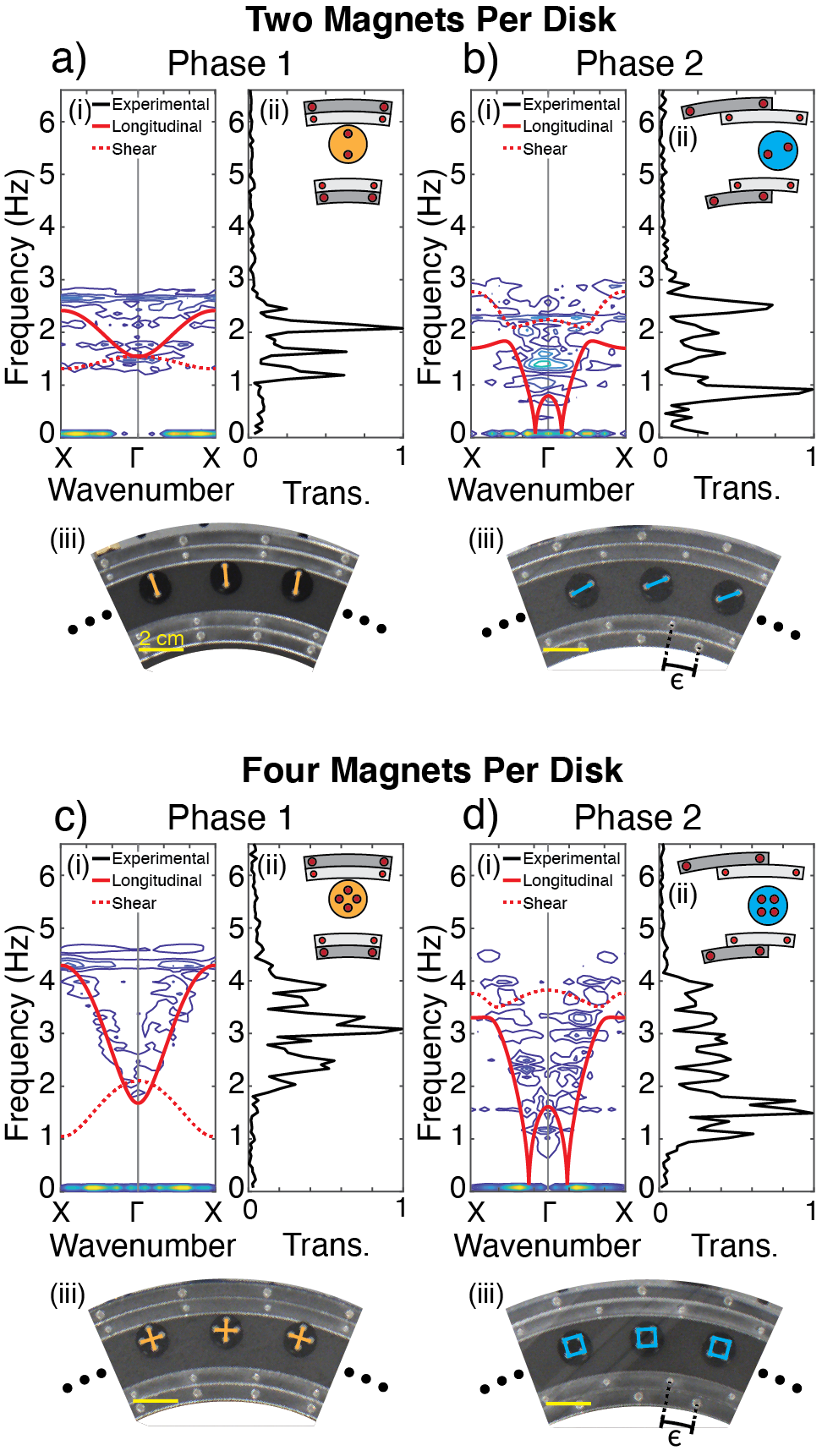}
\caption{\label{fig:circle_exp_disp_fft_updated} \textbf{Transmission through circular channel.} Analytical dispersion curves and experimental transmission ranges for two magnet per disk in a) phase 1 and b) phase 2 orientation. Four magnet per disk behavior in the c) phase 1 and d) phase 2 orientation. (a-d)(i) The analytical (red) and experimental (contour) dispersion curves, (ii) the normalized transmission of a the tenth disk from the source and the unit cell (inset), and (iii) a photo of three unit cells within the metamaterial of the circular channel boundary in the phase 1 or phase 2 orientation (i.e., shifted $0^\circ$ or $7.5^\circ$, respectively).}
\end{center}
\end{figure}

To experimentally validate the self-alignment of the disks within the circular boundary into two topologically distinct phases, we cut concentric circular boundary walls out of acrylic glass sheets resembling those in figure \ref{fig:circle_disk_location_tracking} a, b. The outer diameter of the outermost ring is 300 $mm$, while the inner diameter is 285 $mm$. The thickness of the second largest ring is 7.5 $mm$. The inner diameter of the innermost ring is 190 $mm$, while the outer diameter is 205 $mm$. The thickness of the second smallest ring is 7.5 $mm$. The embedded magnets in the inner and outer sets of boundary walls resemble those in the straight channel boundary (i.e., smaller magnets embedded in the inner set, larger magnets embedded in the outer set) all with the north pole facing upwards. Each unit cell is fabricated in $15^\circ$ increments, totalling 24 unit cells within the circle. We use the same method to fabricate and prepare the disks that were used in the straight channel boundary. We place the boundary on the air bearing table and the disks within the boundary walls in random orientations. We activate the air bearing, creating a layer of laminar flow beneath the disks. The disks self-align into their lowest potential energy locations and orientations. Based on the relative angle between the boundary walls, the disks align in the phase 1 or phase 2 orientations (Fig. \ref{fig:circle_exp_disp_fft_updated} (a-d)(iii)).
\medskip

\begin{figure}
\begin{center}
\includegraphics[scale=1]{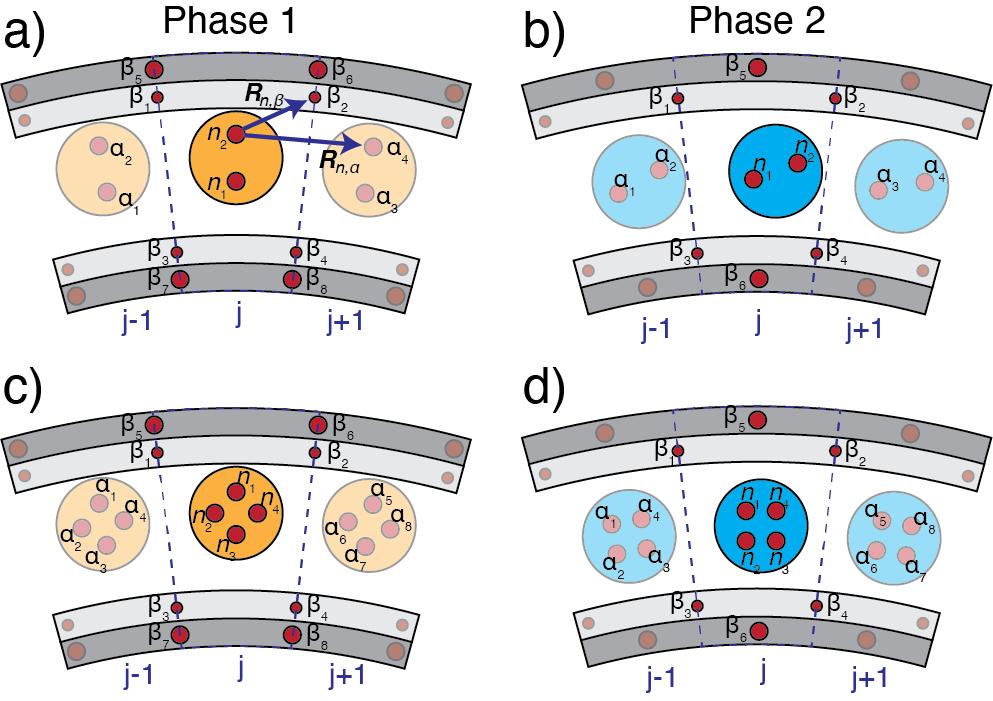}
\end{center}
\caption{\textbf{Circular channel force unit vectors.} Labeling and representation of the unit force vectors within a unit cell, $j$ (opaque). a) Example unit vector between disk $j$ magnets ($n$), boundary magnets ($\beta$), and the adjacent disks' embedded magnets $\alpha$ (transparent) used in the stiffness matrix. Labeling and indexing of two magnets per disk b) phase 2 and four magnets per disk c) phase 1 and d) phase 2.}
\label{fig:supp_circular_channel_vectors}
\end{figure}

Similar to the analysis of the unit cell within the straight boundary, the dispersion analysis for the unit cell within the circular boundary considers the nearest neighbor interactions (Fig. \ref{fig:supp_circular_channel_vectors}). The labeling and indexing is the same for the circular boundary as that for the straight boundary (Fig. \ref{fig:supp_straight_channel_vectors}). The unit cells in the phase 1 orientation (Fig. \ref{fig:supp_circular_channel_vectors} a, c) have eight boundary magnets ($\Gamma=8$), while the unit cells in the phase 2 orientation (Fig. \ref{fig:supp_circular_channel_vectors} b, d) have six boundary magnets ($\Gamma=6$). An example unit vector between disk magnet $n_i$, where $i = 2$, and boundary magnet $\beta_{ii}$, where $ii = 2$ and the same disk magnet $n_{2}$ and adjacent disk magnet $\alpha_{iii}$, where $iii = 4$ is shown in figure \ref{fig:supp_circular_channel_vectors} a. While the dispersion model is similar to that of the straight boundary, in the circular case, the separation distances between the top and the bottom boundaries are different, taking into account the curvature of the unit cell.
\medskip

The analytically computed dispersion curves for the circular boundary are plotted in figure \ref{fig:circle_exp_disp_fft_updated} (a-d)(i). Similar to the straight boundary (Fig. \ref{fig:channel_exp_disp_fft}), the shift from phase 1 to phase 2 has a measurable effect on the lower frequencies of the dispersion curves.  In the phase 1 orientation, for both types of disks, neither of the transmission bands exist below 1 Hz (Fig. \ref{fig:circle_exp_disp_fft_updated} a(i), c(i)). In the phase 2 orientation,  for both types of disks, both transmission bands exist at or below 0.15 Hz (Fig. \ref{fig:circle_exp_disp_fft_updated} b(i), d(i)). This phenomenon is consistent with that seen in the straight boundary case (i.e., the adjustment of the boundary walls and the consequential change in locations and orientations of the floating disks from phase 1 to phase 2 causes lower expected transmission frequencies). In addition, increasing the number of embedded magnets in each disk has a measurable effect on the transmission frequencies. The calculated transmission band (i.e., the dispersion curve frequency range) in the case of two magnets per disk in phase 1 (Fig. \ref{fig:circle_exp_disp_fft_updated} a(i)) spans 1.1 Hz with an upper frequency cutoff of 2.4 Hz. In contrast, the calculated transmission band the case of four magnets per disk in phase 1 (Fig. \ref{fig:circle_exp_disp_fft_updated} c(i)) spans 3.3 Hz (3 times wider) with an upper frequency cutoff of 4.3 Hz.
\medskip

To experimentally verify the analytically computed dispersion curves, we excite the metamaterials with a tangential chirp signal. Throughout the duration of the chirp signal, we capture the motion of each floating disk using a computer vision camera. We post-process the captured images using the DICe software to extract the displacement of each disk over time. We analyze the displacements of the floating disks in each of the four metamaterial phases (characterized separately) using a 1D- and 2D-FFT. The 1D-FFT is the measured transmission through the lattice at the tenth disk from the wave source (Fig. \ref{fig:circle_exp_disp_fft_updated} (a-d)(ii)). The 2D-FFT is the experimental dispersion curve, plotted as a contour map, which we superimpose on the analytical dispersion curve for each metamaterial phase (Fig. \ref{fig:circle_exp_disp_fft_updated} (a-d)(i)). The individual disk transmissions (1D-FFT)  and the post-processed dispersion curves (2D-FFT) match well. Similar to the straight boundary case, for two magnets per disk in the phase 2 orientation, we note the emergence of shear motion, even with only longitudinal excitation present, due to the coupling of the shear and longitudinal modes.
\medskip


\end{appendices}

\end{document}